\documentclass[]{revtex4}

\usepackage{amsmath}
\usepackage{amssymb}
\usepackage{epsfig}
\usepackage{graphics}

\begin{document}

\title{Stabilizing qubit coherence via tracking-control}
\author{Daniel A. Lidar and Sara Schneider\footnote{Current address:
    Atel Trading AG, Switzerland}}
\affiliation{Chemical Physics Theory Group, Chemistry Department, and Center for Quantum
Information and Quantum Control, University of Toronto, 80 St. George
Street, Toronto, Ontario M5S 3H6, Canada }

\begin{abstract}
We consider the problem of stabilizing the coherence of a single qubit
subject to Markovian decoherence, via the application of a control
Hamiltonian, without any additional resources. In this case neither quantum
error correction/avoidance, nor dynamical decoupling applies. We show that
using tracking-control, i.e., the conditioning of the control field on the
state of the qubit, it is possible to maintain coherence for finite time
durations, until the control field diverges.
\end{abstract}

\maketitle

\section{Introduction}

Protecting quantum coherence in the presence of decoherence due to coupling
to an uncontrollable environment is an important goal of quantum control,
with applications in, e.g., quantum information processing \cite{Nielsen:book}, and coherent control \cite{Brumer:book}. Various methods
have been developed for this purpose, e.g., quantum error correcting codes 
\cite{Steane:99}, decoherence-free subspaces \cite{LidarWhaley:03},
and dynamical decoupling \cite{Viola:01a}, and combinations thereof 
\cite{ByrdWuLidar:review}. However, none of these methods is applicable in
the simplest possible case of interest, of a \emph{single} qubit subject to 
\emph{Markovian} decoherence: both quantum error correcting codes and
decoherence-free subspaces rely on an encoding of the state of the qubit
into the state of \emph{several} qubits, whereas dynamical decoupling is 
\emph{inapplicable} in the fully Markovian regime, since it is effectively
equivalent to the quantum Zeno effect \cite{Facchi:03}, i.e., requires in an
essential manner that the bath retains some memory of its interaction
with the system.

In this work we show that \emph{tracking-control} is capable of stabilizing the 
\emph{coherence} of a single qubit subject to Markovian decoherence.
Tracking-control has a rich history in classical control theory
\cite{Hirschorn:79,Hirschorn:81,Hirschorn:88,Jakubczyk:93,Retchkiman:95}.
It was extensively studied by
Rabitz and co-workers in the context of closed quantum systems
(specifically, molecular systems)
\cite{Chen:95,Lu:95,Gross:93,Zhu:99,Zhu:98,Zhu:03}. Here it
refers to the instantaneous adjustment of the control field based on a
continuous measurement of the state of the qubit. This is, of course, a very
strong assumption, which may be appropriate for classical control, but
cannot be satisfied in principle if non-commuting quantum observables are
involved. It turns out that we are able to overcome this obstacle by
choosing a specific form for our control fields, as explained in detail
below. In general, one can envision performing quantum tracking control
based on the incomplete information gleaned in real-time from measuring an
as large as possible set of commuting observables. It is an open problem to
estimate the quality of the (partial) tracking one can thus attain.

The tracking solution we find in our case becomes singular after a finite
time (i.e., the control fields diverge). Such singularities are a well-known
feature of tracking control \cite{Chen:95,Lu:95,Gross:93,Zhu:99}, and can in
some cases be removed \cite{Zhu:99}. We analyze the nature of the
singularity occurring in our case, finding that is is unavoidable.

A related control problem was recently addressed by Belavkin and co-workers
in Ref.~\cite{Bouten:04}, using the quantum filtering (or stochastic
master) equation \cite{Belavkin:80}. In
their work the control objective is to take a qubit from an unknown initial
state to the $|0\rangle $ state. In this approach one naturally accounts for
the unavailability of complete information from the measurement process
(using filtering), but the resulting quantum Bellman-Hamilton-Jacobi
equation is very hard to solve even for a qubit, unless one makes drastic
simplifying assumptions about the control field.

Purely unitary control in the presence of
decoherence has also been addressed by Recht {\it et al}. in Ref.~\cite{Recht:02}. It was shown there that in the case of
so-called relaxing semigroups (quantum dynamical semigroups \cite%
{Alicki:87} with a unique fixed point), one can control the equilibrium
state of the dynamics. Relaxing semigroups are in fact a case that is
orthogonal to the case we study here, as they involve non-unital channels,
while we study only unital channels. These issues are clarified below.

Finally, the problem of coherent control of a qubit subject to Markovian
dynamics has been studied in detail by Altafini, using a Lie algebraic
framework \cite{Altafini:03,Altafini:04}. This work has elucidated the
corresponding conditions for accessibility, small- and finite-time
controllability, using the coherence vector representation \cite{Alicki:87}
and employing classical control-theory notions.

The structure of the present paper is as follows. In the next Section we
define the model of a single qubit subject to Markovian decoherence, and
define our control objectives in terms of the purity and coherence. In Sec.~\ref{tracking} we set up the coherence tracking problem and solve it
explicitly for a qubit subject to pure dephasing. In Sec.~\ref{equiv} we
discuss the generality of this result by defining equivalence classes of the
pure dephasing channel to which our coherence-tracking solution also
applies. In Sec.~\ref{singularity} we discuss the nature of the singularity
of our control fields, and show that this singularity cannot be avoided. We
conclude in Sec.~\ref{conclusions} with a brief summary and a list of open
questions inspired by this work. 

\section{Model and Objectives}

\subsection{Purity}

\label{purity}

We consider a single qubit subject to Markovian decoherence and controlled
via a control Hamiltonian $H$. Then the system dynamics is governed by a
master equation of the form 
\begin{equation}
\frac{\partial \rho }{\partial t}=-\frac{i}{\hbar }[H,\rho ]+{\mathcal{L}}%
(\rho ),  \label{eq:Lind}
\end{equation}%
where the Lindbladian is%
\begin{equation}
{\mathcal{L}}(\rho )=\frac{1}{2}\sum_{i,j}a_{ij}\left( [F_{i},\rho
F_{j}^{\dagger }]+[F_{i}\rho ,F_{j}^{\dagger }]\right) ,  \label{lind}
\end{equation}%
where the matrix $A=(a_{ij})$ is positive semidefinite [ensuring completely
positivity of the mapping ${\mathcal{L}}(\rho )$], and where the Lindblad
operators $\{F_{i}\}$ are the coupling operators of the system to the bath 
\cite{Alicki:87}. One can always diagonalize $A$ using a unitary transformation 
$W=(w_{ij})$ and define new Lindblad operators $G_{i}=\sum w_{ij}F_{j}$ such
that 
\begin{equation}
{\mathcal{L}}(\rho )=\frac{1}{2}\sum_{i}\gamma _{i}\left( [G_{i},\rho
G_{i}^{\dagger }]+[G_{i}\rho ,G_{i}^{\dagger }]\right) ,  \label{eq:lind2}
\end{equation}%
where $\gamma _{i}\geq 0$ are the eigenvalues of $A$.

In such a Markovian system \emph{it is impossible to control the purity} 
\begin{equation}
p\equiv \mathrm{Tr}(\rho ^{2})
\end{equation}%
of a state $\rho $ in such a way so as to maintain it at its initial value 
\cite{Ketterle:92,Tannor:99}. To see this note that the time derivative of $%
p $ is given by 
\begin{eqnarray}
\frac{\partial p}{\partial t} &=&\frac{\partial \mathrm{Tr}(\rho ^{2})}{%
\partial t}=2\mathrm{Tr}(\rho \dot{\rho})  \notag \\
&=&-\frac{2i}{\hbar }\underbrace{\mathrm{Tr}(\rho \lbrack H,\rho ])}_{=0}+2%
\mathrm{Tr}(\rho {\mathcal{L}}(\rho )),
\end{eqnarray}%
where we used cyclic invariance under the trace. Thus the Hamiltonian
control term cannot change the first derivative of the purity, and hence
cannot keep it at its initial value. This is also known as the
\textquotedblleft no-cooling principle\textquotedblright .

Moreover, for certain Lindbladians purity is a strictly decreasing function
under the Markovian semigroup dynamics. This is the case for all unital
Lindbladians, i.e., those for which ${\mathcal{L}}(I)=0$ (e.g., \cite{Altafini:04}). In this case one finds from Eq.~(\ref{eq:lind2}) that $%
\sum_{i}\gamma _{i}\left( G_{i}G_{i}^{\dagger }-G_{i}^{\dagger }G_{i}\right)
=0$. Thus a sufficient condition for unitality is that the $G_{i}$ are \emph{%
normal} (e.g., unitary):\ $G_{i}G_{i}^{\dagger }=G_{i}^{\dagger }G_{i}$.
Subject to normality it is possible give a simple proof of the monotonic
decrease of $p$:%
\begin{eqnarray*}
\frac{\partial p}{\partial t} &=&2\mathrm{Tr}(\rho {\mathcal{L}}(\rho )) \\
&=&\sum_{i}\gamma _{i}\mathrm{Tr}\{\rho \left( \lbrack G_{i},\rho
G_{i}^{\dagger }]+[G_{i}\rho ,G_{i}^{\dagger }]\right) \} \\
&=&2\sum_{i}\gamma _{i}\{\mathrm{Tr}(X_{i}Y_{i})-\frac{1}{2}[\mathrm{Tr}%
(X_{i}X_{i}^{\dagger })+\mathrm{Tr}(Y_{i}Y_{i}^{\dagger })]\},
\end{eqnarray*}%
where $X_{i}=\rho G_{i}$ and $Y_{i}=\rho G_{i}^{\dagger }$, and we used $%
\rho =\rho ^{\dag }$. Now apply the arithmetic-geometric mean inequality for
matrices, $\left\Vert XY\right\Vert \leq \frac{1}{2}(\left\Vert X^{\dagger
}X+YY^{\dagger }\right\Vert )$, where $\left\Vert \cdot \right\Vert $ is any
unitarily-invariant norm (such as $\mathrm{Tr}$) \cite{Bhatia:book}[p.263].
Also, $\left\Vert X^{\dagger }X+YY^{\dagger }\right\Vert \leq \left\Vert
X^{\dagger }X\right\Vert +\left\Vert YY^{\dagger }\right\Vert $. Then:%
\begin{eqnarray}
\mathrm{Tr}(X_{i}Y_{i})-\frac{1}{2}[\mathrm{Tr}(X_{i}X_{i}^{\dagger })+%
\mathrm{Tr}(Y_{i}Y_{i}^{\dagger })] &\leq &\frac{1}{2}[\mathrm{Tr}%
(X_{i}^{\dagger }X_{i}+Y_{i}Y_{i}^{\dagger })]-\frac{1}{2}[\mathrm{Tr}%
(X_{i}X_{i}^{\dagger })+\mathrm{Tr}(Y_{i}Y_{i}^{\dagger })]  \notag \\
&\leq &0.
\end{eqnarray}%
Thus, using $\gamma _{i}\geq 0$, we have $\frac{\partial p}{\partial t}\leq 0
$. Purity in the case of unital Lindbladians is, therefore, a strictly
decaying function under Hamiltonian control. This conclusion is unchanged
even with feedback, namely, even if the Hamiltonian includes dependence on
the qubit state, i.e., if $H=H[\rho ]$. The situation is different for
non-unital Lindbladians \cite{Recht:02,Altafini:04}. These channels can increase the
purity even without active control. E.g., under spontaneous emission an
arbitrary qubit mixed-state is gradually purified to $|0\rangle $. In this
work we consider only unital Lindbladians.

\subsection{Coherence}

Another quantum quantity of relevance is \emph{coherence}, i.e., the
off-diagonal elements of $\rho $. This is, of course, a basis-dependent
quantity (it is not invariant under unitary transformations), so that in
what follows we assume that one has fixed a basis for physical reasons
(e.g., there is a magnetic field pointing in the $z$ direction, or there is
a pure dephasing type coupling to the bath, represented by a $\sigma _{z}$%
-Lindblad operator). We will show that tracking control is capable of
stabilizing the coherence of a single qubit. A qubit is completely
characterized by its density matrix 
\begin{equation}
\rho =\left( 
\begin{array}{cc}
\rho _{00} & \rho _{01} \\ 
\rho _{10} & \rho _{11}%
\end{array}%
\right) ,
\end{equation}%
with the additional constraints $\mathrm{Tr}(\rho )=1$, $\rho _{10}=\rho
_{01}^{\ast }$ and $\mathrm{Tr}(\rho ^{2})\leq 1$. We follow the approach in 
\cite{Alicki:87} and parametrize $\rho $ using the Bloch
vector $\vec{v}$ with the \emph{real} components 
\begin{equation}
v_{\alpha }=\frac{1}{2}\mathrm{Tr}(\rho \sigma _{\alpha }),
\end{equation}%
where $\sigma _{\alpha }$, $\alpha \in \{x,y,z\}$, are the Pauli matrices,
and we identify the Lindblad operators as $F_{1}=\sigma _{x},F_{2}=\sigma
_{y},F_{3}=\sigma _{z}$. We then have 
\begin{equation}
\rho =\frac{1}{2}{(I}+\vec{v}\cdot \vec{\sigma}).  \label{eq:rho-v}
\end{equation}%
The purity $p$ is then given by the \emph{Bloch sphere radius}, 
\begin{equation}
p=v_{x}^{2}+v_{y}^{2}+v_{z}^{2},
\end{equation}%
while the coherence $c$ between levels $|0\rangle $ and $|1\rangle $ is
given by the \emph{radius in the }$x-y$\emph{\ plane}, 
\begin{equation}
c=v_{x}^{2}+v_{y}^{2}.
\end{equation}

\emph{Our objective is to have }$c$\emph{\ be constant during the evolution}%
. Thus we impose the following constraint:%
\begin{eqnarray}
\frac{1}{2}\frac{\partial c}{\partial t} &=&v_{x}\dot{v}_{x}+v_{y}\dot{v}_{y}
\notag \\
&=&0.  \label{cohconst}
\end{eqnarray}

\section{Tracking Control of Coherence}

\label{tracking}

\subsection{Tracking equation}

Noticing that the traceful part (reference energy) of the control
Hamiltonian drops out of the commutator $[H,\rho ]$ we expand $H$ in the
traceless Pauli basis: 
\begin{equation}
H=\frac{\hbar }{2}(\omega _{0}(t)\sigma _{z}+\omega _{1}(t)\sigma
_{x}-\omega _{2}(t)\sigma _{y})
\end{equation}%
We wish to solve for the control fields $\omega _{i}(t)$ so that the
coherence constraint (\ref{cohconst}) is satisfied. Note that we are
assuming that the intrinsic system Hamiltonian either vanishes, or that we
are working in the rotating frame with respect to this Hamiltonian.

Substituting this and the expansion (\ref{eq:rho-v}) into the Lindblad
equation (\ref{eq:Lind}), and using trace-orthonormality of the Pauli
matrices, we obtain the generalized Bloch equations 
\begin{eqnarray}
\dot{v}_{x}(t) &=&-\gamma _{3}v_{x}(t)+(\alpha -\omega
_{0}(t))v_{y}(t)+(\beta -\omega _{2}(t))v_{z}(t)-2\lambda ,  \label{vx} \\
\dot{v}_{y}(t) &=&(\alpha +\omega _{0}(t))v_{x}(t)-\gamma
_{2}v_{y}(t)+(\delta -\omega _{1}(t))v_{z}(t)-2\mu ,  \label{vy} \\
\dot{v}_{z}(t) &=&(\beta +\omega _{2}(t))v_{x}(t)+(\delta +\omega
_{1}(t))v_{y}(t)-\gamma _{1}v_{z}(t)-2\nu .  \label{vz}
\end{eqnarray}%
Here we identify the parameters from Eqs.~(\ref{vx})-(\ref{vz}) with the $%
a_{ij}$ from Eq.~(\ref{lind}) as follows: 
\begin{eqnarray}
\gamma _{1} &=&2(a_{22}+a_{33}),\quad \gamma _{2}=2(a_{11}+a_{33}),\quad
\gamma _{3}=2(a_{11}+a_{22}) \\
\alpha &=&2\mathrm{Re}(a_{12}),\quad \beta =2\mathrm{Re}(a_{13}),\quad
\delta =2\mathrm{Re}(a_{23}) \\
\lambda &=&\mathrm{Im}(a_{23}),\quad \mu =-\mathrm{Im}(a_{13}),\quad \nu =%
\mathrm{Im}(a_{12})
\end{eqnarray}%
Note from Eqs.~(\ref{vx}-\ref{vz}) that the $\gamma _{i}$ can be interpreted
as damping coefficients, while $\alpha ,\beta $ and $\delta $ play the role
of Lamb shifts (modify the control fields), and $\lambda ,\mu $ and $\nu $
are the coordinates of an affine shift of the Bloch vector (which plays a
role, e.g., in spontaneous emission). Positive semi-definiteness of $%
A=(a_{ij})$ imposes various conditions on these parameters \cite{Alicki:87}.

In matrix form Eqs.~(\ref{vx}-\ref{vz}) can be written as an affine linear
transformation 
\begin{equation}
\dot{\vec{v}}(t)=(M_{0}+M(t))\vec{v}(t)+\vec{k},  \label{eq:vdot1}
\end{equation}%
where the decoherence is affected by 
\begin{eqnarray}
M_{0} &=&\left( 
\begin{array}{ccc}
-\gamma _{3} & \alpha & \beta \\ 
\alpha & -\gamma _{2} & \delta \\ 
\beta & \delta & -\gamma _{1}%
\end{array}%
\right) ,  \notag \\
\vec{k} &=&-2(\lambda ,\mu ,\nu )^{t},  \label{eq:M0k}
\end{eqnarray}%
(superscript $t$ denotes transpose) and the control Hamiltonian is represented by the real, antisymmetric matrix%
\begin{eqnarray}
M(t) &=&\left( 
\begin{array}{ccc}
0 & -\omega _{0}(t) & -\omega _{2}(t) \\ 
\omega _{0}(t) & 0 & -\omega _{1}(t) \\ 
\omega _{2}(t) & \omega _{1}(t) & 0%
\end{array}%
\right)  \notag \\
&=&\sum_{j=0}^{2}\omega _{j}(t)\Lambda _{j},  \label{eq:M(t)}
\end{eqnarray}%
where the matrices%
\begin{equation}
\Lambda _{0}=\left( 
\begin{array}{ccc}
0 & -1 & 0 \\ 
1 & 0 & 0 \\ 
0 & 0 & 0%
\end{array}%
\right) ,\quad \Lambda _{1}=\left( 
\begin{array}{ccc}
0 & 0 & 0 \\ 
0 & 0 & -1 \\ 
0 & 1 & 0%
\end{array}%
\right) ,\quad \Lambda _{2}=\left( 
\begin{array}{ccc}
0 & 0 & -1 \\ 
0 & 0 & 0 \\ 
1 & 0 & 0%
\end{array}%
\right)  \label{eq:lambdas}
\end{equation}%
close as an $\mathrm{so}(3)$ subalgebra of $\mathrm{su}(3)$ (we use a
slightly different convention for the signs than the standard one \cite{Cornwell:84II}):
\begin{equation}
\lbrack \Lambda _{0},\Lambda _{1}]=-\Lambda _{2},\quad \lbrack \Lambda
_{1},\Lambda _{2}]=-\Lambda _{0},\quad \lbrack \Lambda _{2},\Lambda
_{0}]=-\Lambda _{1}.
\end{equation}

Solving Eq.~(\ref{vy}) for $\omega _{1}(t)$ and Eq.~(\ref{vx}) for $\omega
_{2}(t)$ we obtain 
\begin{eqnarray}
\omega _{1}(t) &=&\frac{(\alpha +\omega _{0}(t))v_{x}(t)-\gamma
_{2}v_{y}(t)+\delta v_{z}(t)-2\mu -\dot{v}_{y}(t)}{v_{z}(t)},  \label{om1} \\
\omega _{2}(t) &=&\frac{-\gamma _{3}v_{x}(t)+(\alpha -\omega
_{0}(t))v_{y}(t)+\beta v_{z}(t)-2\lambda -\dot{v}_{x}(t)}{v_{z}(t)}.
\label{om2}
\end{eqnarray}%
Substituting this into Eq.~(\ref{vz}) and using the imposed constraint of
constant coherence, $\dot{c}=0$, we find a non-linear first-order
differential equation for $v_{z}(t)$ given by 
\begin{equation}
\dot{v}_{z}(t)=F(t)+\frac{G(t)}{v_{z}(t)}-\gamma _{1}v_{z}(t),
\label{eq:track}
\end{equation}%
where we have defined 
\begin{eqnarray}
F(t) &\equiv &2\beta v_{x}(t)+2\delta v_{y}(t)-2\nu ,  \label{eq:F} \\
G(t) &\equiv &-\gamma _{3}v_{x}^{2}(t)-\gamma _{2}v_{y}^{2}(t)+2\alpha
v_{x}(t)v_{y}(t)-2\lambda v_{x}(t)-2\mu v_{y}(t).  \label{eq:G}
\end{eqnarray}%
Eq.~(\ref{eq:track}) is our tracking equation.

\subsection{Solution of the tracking equation in the case of pure dephasing}

We first consider the relatively simple case of diagonal $A$ (which implies
that $\alpha =\beta =\delta =\lambda =\mu =\nu =0$), with equal damping
along the $x$ and $y$ axes ($\gamma _{2}=\gamma _{3}\equiv \gamma $), and $%
\gamma _{1}=0$. This is the case known as pure dephasing (see below), and is
generalized in Section~\ref{equiv}. The Bloch equations then become:%
\begin{eqnarray}
\frac{\partial \vec{v}}{\partial t} &=&\left( 
\begin{array}{ccc}
-\gamma & -\omega _{0}(t) & -\omega _{2}(t) \\ 
\omega _{0}(t) & -\gamma & -\omega _{1}(t) \\ 
\omega _{2}(t) & \omega _{1}(t) & 0%
\end{array}%
\right) \vec{v}  \notag \\
&=&\vec{\Omega}(t)\times \vec{v}(t)+\vec{\Gamma}\cdot \vec{v}(t),
\label{eq:Bloch-simp}
\end{eqnarray}%
where%
\begin{eqnarray}
\vec{\Omega}(t) &=&(\omega _{1}(t),-\omega _{2}(t),\omega _{0}(t)),  \notag
\\
\vec{\Gamma} &=&(-\gamma ,-\gamma ,0).
\end{eqnarray}%
The vector $\vec{\Omega}$ acts as an effective time-dependent magnetic field
and rotates the coherence vector in a manner designed (see below) to keep
the coherence constant for as long as possible.

\subsubsection{Uncontrolled (free) dynamics}

Under free evolution (uncontrolled scenario: $H=0$) the system dynamics is
governed by Markovian decoherence, subject to a master equation of the form 
\begin{equation}
\frac{\partial \rho }{\partial t}=-\frac{i}{\hbar }[H,\rho ]+\frac{\gamma }{2%
}\left( \sigma _{z}\rho \sigma _{z}-\rho \right) ,  \label{eq:mastereq}
\end{equation}%
or, equivalently, to the following Bloch equations [Eq.~(\ref{eq:Bloch-simp}%
)]: 
\begin{equation}
\dot{v}_{x}(t)=-\gamma v_{x}(t),~\dot{v}_{y}(t)=-\gamma v_{y}(t),~\dot{v}%
_{z}(t)=0.
\end{equation}%
The solution is 
\begin{eqnarray}
v_{x}(t) &=&\exp [-\gamma t]v_{x}(0),  \notag \\
v_{y}(t) &=&\exp [-\gamma t]v_{y}(0),  \notag \\
v_{z}(t) &=&v_{z}(0).  \label{eq:free}
\end{eqnarray}%
This is known as pure dephasing, or a phase-flip channel \cite{Nielsen:book}%
, since in the corresponding Kraus operator-sum representation 
\begin{eqnarray}
\rho (t) &=&\left( 
\begin{array}{cc}
\rho _{00} & e^{-\gamma t}\rho _{01} \\ 
e^{-\gamma t}\rho _{01}^{\ast } & 1-\rho _{00}%
\end{array}%
\right)  \notag \\
&=&(1-a)I\rho I+a\sigma _{z}\rho \sigma _{z},
\end{eqnarray}%
the qubit undergoes a phase flip with probability $a=(1-e^{-\gamma t})/2$ ($%
I $ is the identity matrix).

\subsubsection{Controlled dynamics}

Eqs.~(\ref{eq:track})-(\ref{eq:G}) simplify considerably in the pure
dephasing case, and using the expression $c=v_{x}^{2}+v_{y}^{2}$ (is
constant) for the coherence we find from Eq.~(\ref{eq:track}): 
\begin{equation}
\dot{v}_{z}(t)=\frac{-\gamma c}{v_{z}(t)}.
\end{equation}%
Multiplying by $v_{z}(t)$ and integrating $\frac{1}{2}d(v_{z}(t))^{2}/dt$,
the solution is 
\begin{equation}
v_{z}(t)=\left( \pm \right) \sqrt{-2\gamma ct+v_{z}^{2}(0)}.
\label{eq:vz(t)}
\end{equation}%
It is clear that $v_{z}(t)$ does not stay real for times $t>t_{b}$ where 
\begin{equation}
t_{b}=\frac{v_{z}^{2}(0)}{2\gamma c}.  \label{eq:tb}
\end{equation}%
What happens is that $v_{z}(t_{b})=0$, so that purity equals coherence.
Since the control fields trade decrease in purity in return for
stabilization of coherence, at the breakdown point this trade-off becomes
impossible. Mathematically, the constraint $\dot{c}=0$ forces the
control fields to diverge. We further investigate the implications of this
breakdown below.

We can solve for the corresponding control fields from Eqs.~(\ref{om1}),(\ref{om2}), yielding 
\begin{eqnarray}
\omega _{2}(t) &=&\left( \pm \right) \frac{-\gamma v_{x}(t)-\omega
_{0}(t)v_{y}(t)-\dot{v}_{x}}{\sqrt{-2\gamma ct+v_{z}^{2}(0)}},  \label{cont1}
\\
\omega _{1}(t) &=&\left( \pm \right) \frac{\omega _{0}(t)v_{x}(t)-\gamma
v_{y}(t)-\dot{v}_{y}(t)}{\sqrt{-2\gamma ct+v_{z}^{2}(0)}}.  \label{cont2}
\end{eqnarray}%
Note that the field $\omega _{0}(t)$ can be chosen arbitrarily. Also note
that the fields $\omega _{1}(t),\omega _{2}(t)$ depend on $v_{x}(t)$ and $%
v_{y}(t)$. This is why the method is called \textquotedblleft tracking
control\textquotedblright :\ the control strategy depends on the
instantaneous state of the system we desire to control. This can be
technically highly demanding, since it implies the ability to make very fast
measurements (i.e., much faster than the decoherence time-scale), combined
with classical processing to solve for the control fields and real-time
feedback. Nevertheless, recent cavity-QED experiments haven demonstrated the
possibility of such real-time feedback control \cite{Steck:04}. For a
quantum-mechanical system such as a qubit, tracking control involves an
intrinsically undesirable feature: simultaneous knowledge of $v_{x}(t)$ and $%
v_{y}(t)$ is impossible since $\sigma _{x}$ and $\sigma _{y}$ are
non-commuting observables. However, there is a simple fix for this problem,
once we realize that the time dependence of $v_{x}(t)$ and $v_{y}(t)$ is
itself induced~by the control fields. Thus, we can use fields that fix $%
v_{x}(t)=v_{x}(0)$ and $v_{y}(t)=v_{y}(0)$, which is, of course, just a
particular way of keeping the coherence $c=v_{x}^{2}+v_{y}^{2}$ constant, 
\emph{via linearization of the control objective} (we have taken a quadratic
control objective and replaced it by two linear objectives). We then find
that the required fields have the form%
\begin{eqnarray}
\omega _{2}(t) &=&\left( \pm \right) \frac{-\gamma v_{x}(0)-\omega
_{0}(t)v_{y}(0)}{\sqrt{v_{z}^{2}(0)-2\gamma ct}},  \label{eq:om2-fin} \\
\omega _{1}(t) &=&\left( \pm \right) \frac{-\gamma v_{y}(0)+\omega
_{0}(t)v_{x}(0)}{\sqrt{v_{z}^{2}(0)-2\gamma ct}}.  \label{eq:om1-fin}
\end{eqnarray}%
\emph{At the breakdown time }$t_{b}$\emph{\ the control fields diverge}, as
can be seen already from Eqs.~(\ref{cont1}),(\ref{cont2}). Thus, decay of
the coherence can be prevented for $t<t_{b}$.

\subsection{Analysis}

\label{analysis}

The breakdown time $t_{b}=v_{z}^{2}(0)/(2\gamma c)$ is inversely
proportional to the desired constant initial coherence value $c$: the higher
the initial coherence we wish to maintain, the less time this can be done
for. The breakdown time also depends on $v_{z}^{2}(0)=p(0)-c$, where $p$ is
the purity: If $v_{z}(0)=0$ (coherence$=$purity) the coherence will start to
decay immediately and no control is possible. Note that $v_{z}(0)=0$ is the
state of maximum coherence, as can be seen from the condition $p(0)\leq 1$
(the radius in the $x-y$ plane equals the Bloch sphere radius). Thus,
coherence control is, clearly, strongly state-dependent.

\begin{figure}[h]
\centering
\resizebox{7cm}{!}{\includegraphics{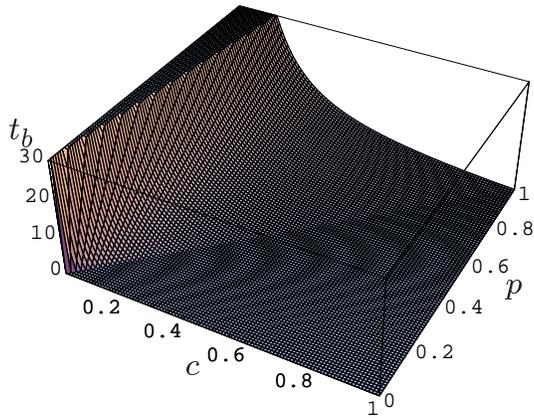}}
\caption{The time for which the coherence of a qubit can be kept constant,
as a function of coherence $c$ and purity $p$, for $\protect\gamma =0.1$.}
\label{fig1}
\end{figure}

In Fig.~\ref{fig1} we plot the breakdown time $t_{b}$ as a function of
coherence $c$ and purity $p$, for $\gamma =0.1$. The plot reflects the
constraint $c\leq p$, and shows the tradeoff between coherence and the time
for which it can be maintained. 

\begin{figure}[h]
\centering
\resizebox{7cm}{!}{\includegraphics{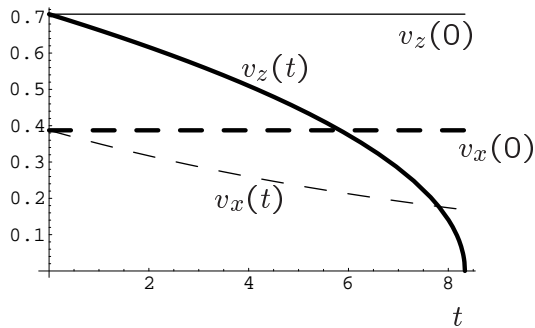}}
\caption{Comparison between the controlled (thick lines) and the free (thin
lines) evolution for $v_{z}$ and $v_{x}$. Here $\protect\gamma =0.1$, $c=0.3$%
, $p=0.8$, $v_{x}(0)=v_{y}(0)$.}
\label{fig2}
\end{figure}

It is interesting to compare the controlled and uncontrolled dynamics. We
again set $\gamma =0.1$, and in the controlled scenario choose an initial
state which lies in a region of Fig.~\ref{fig1} where there is some
coherence to be preserved: $c=0.3$ and $p=0.8$. This implies $v_{z}(0)=\sqrt{%
p(0)-c}=1/\sqrt{2}$. We set $v_{x}(0)=v_{y}(0)$, so that $v_{x}(0)=\sqrt{c/2}%
=\sqrt{0.3/2}\approx 0.39$. A comparison of the two cases is plotted in
Fig.~\ref{fig2}. In the uncontrolled scenario of pure dephasing $v_{z}$ is
a constant of motion, while $v_{x}$ decays monotonically. In contrast, in
the controlled case these roles are reversed in spite of the dephasing:\ the
control fixes $v_{x}$ while $v_{z}$ is allowed to decay.

\begin{figure}[h]
\centering
\resizebox{7cm}{!}{\includegraphics{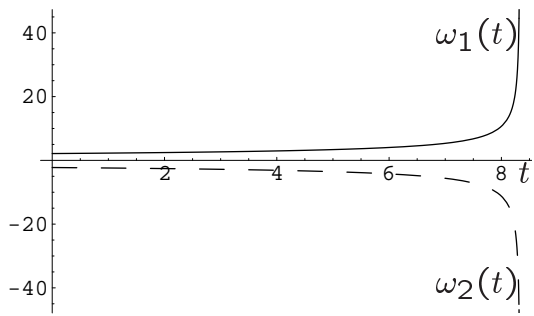}}
\caption{Control fields $\protect\omega _{1}$ (solid line) and $\protect%
\omega _{2}(t)$ (dashed line). Parameters as in Fig.~(\protect\ref{fig2}),
and $\protect\omega _{0}=4.0$.}
\label{fig3}
\end{figure}

The control fields necessary to keep the coherence constant are plotted in
Fig.~(\ref{fig3}), where we have chosen $\omega _{0}(t)\equiv \omega _{0}$
to account for an energy difference between the two qubit states. For the
plot we set $\omega _{0}=4.0$. All other parameters and the initial state
are as in Fig.~(\ref{fig2}). The divergence of the control fields at $t_{b}$
is clearly visible. 

Geometrically, the uncontrolled phase-flip channel maps the the Bloch sphere
to an ellipsoid with the $z$-axis as major axis and minor axis in the $x-y$ plane.
As is clear from Eqs.~(\ref{eq:free}), the major axis is invariant under
the uncontrolled dynamics, while the minor axis (the coherence) is
contracted. On the other hand, in the controlled scenario, the minor axis is
invariant up until the breakdown time ($c$ is constant), while the major
axis is contracted [Eq.~(\ref{eq:vz(t)})]. The control field is thus able to
trade the contraction in the $x-y$ plane for one along the $z$-axis. The
geometric interpretation of this process is the following:\ \emph{the
control field attempts to rotate the ellipsoid so that the minor axis
becomes as aligned as possible with the }$z$\emph{-axis, where it would
experience no contraction}. The rotation takes the minor axis to an
invariant point, which requires ever-growing control field amplitude, until
the contraction is so strong that the control field is no longer capable of
sustaining the required rotation, and diverges. The effectiveness of this
process depends on the desired value of coherence that is to be maintained,
and the initial purity. This rotational interpretation follows from Eq.~(\ref%
{eq:Bloch-simp}):\ we have, in the simplest case of constant $\omega _{0}(t)$%
, $\vec{\Omega}(t)=(\omega _{1}(t),-\omega _{2}(t),\omega _{0})$ [a force
vector whose length in the $x-y$ plane is growing monotonically for $0\leq
t\leq t_{b}$ ($|\vec{\Omega}(t)|^{2}=\frac{(\gamma ^{2}+\omega _{0}^{2})c}{%
v_{z}^{2}(0)-2\gamma ct}+\omega _{0}^{2}$)], multiplying via the vector
cross-product the vector $\vec{v}(t)=(v_{x}(0),v_{y}(0),\sqrt{%
v_{z}^{2}(0)-2\gamma ct})$ [a vector with fixed coordinates in the $x-y$
plane whose magnitude is shrinking monotonically], thus producing a rate of
change of $\vec{v}(t)$ pointing in the plane orthogonal to $\vec{\Omega}(t)$
and $\vec{v}(t)$.

\section{Decoherence equivalence classes}

\label{equiv}

From the geometric interpretation given above it is clear that there are
other decoherence models where the Bloch sphere experiences a similar
deformation, but the contraction happens in a different plane. Then the
question arises which decoherence models are \textquotedblleft
equivalent\textquotedblright\ in the sense that what we have learned from
the above phase flip channel can be applied directly to another decoherence
model.

Suppose we apply a global unitary transformation $U$ to each Lindblad
operator in the master equation (\ref{eq:Lind}), as well as to the control
Hamiltonian $H$. \emph{We will say that two decoherence problems which are
thus related are in the same unitary equivalence class}. Under this
transformation the master equation becomes%
\begin{equation}
\frac{\partial \rho }{\partial t}=-\frac{i}{\hbar }[UHU^{\dagger },\rho ]+%
\frac{1}{2}\sum_{i,j}a_{ij}\left( [UF_{i}U^{\dagger }\rho ,UF_{j}^{\dagger
}U^{\dagger }]+[UF_{i}U^{\dagger },\rho UF_{j}^{\dagger }U^{\dagger
}]\right) .
\end{equation}%
This is the \textquotedblleft Heisenberg picture\textquotedblright . We can
transform to the \textquotedblleft Schrodinger picture\textquotedblright\ by
multiplying the latter master equation by $U$ from the right and $U^{\dagger
}$ from the left: 
\begin{equation}
\frac{\partial \rho ^{\prime }}{\partial t}=-\frac{i}{\hbar }[H,\rho
^{\prime }]+\frac{1}{2}\sum_{ij}a_{ij}\left( [F_{i}\rho ^{\prime
},F_{j}]+[F_{i},\rho ^{\prime }F_{j}]\right) ,  \label{eq:rho'}
\end{equation}%
where $\rho ^{\prime }=U^{\dagger }\rho U$. This is the same as the original
master equation, but for a transformed $\rho $. E.g., the phase flip channel
($\sigma _{z}$-decoherence) is unitarily related to the bit flip channel ($%
\sigma _{x}$-decoherence) via the Hadamard matrix $U=\frac{1}{\sqrt{2}}%
\left( 
\begin{array}{cc}
1 & 1 \\ 
1 & -1%
\end{array}%
\right) $.

In the coherence vector representation the unitary transformation $U$
becomes a real rotation matrix $R$ of the coherence vector via the
appropriate adjoint representation \cite{Cornwell:84II}; e.g., for the qubit
problem $U\in \mathrm{SU}(2)$ while $R\in \mathrm{SO}(3)$. The analog of
Eq.~(\ref{eq:rho'}) [i.e., the transformed Eq.~(\ref{eq:vdot1})] is%
\begin{equation}
\dot{\vec{v}}^{\prime }=(M_{0}+M(t))\vec{v}^{\prime }+\vec{k}
\label{eq:vdot}
\end{equation}%
where $M_{0}$ and $\vec{k}$ are the effects of decoherence in the coherence
vector representation \cite{Alicki:87}, $\vec{v}^{\prime }=R\vec{v}$, and $%
M(t)$ represents the Hamiltonian control [recall Eq.~(\ref{eq:M(t)})].

Thus, two decoherence problems which are in the same unitary equivalence
class differ, in the \textquotedblleft Schrodinger picture\textquotedblright
, by a fixed rotation of the coherence vector $\vec{v}$. In terms of the
control problem at hand, the stabilization of the coherence $c$, it is then
clear that the control solution for all decoherence problems which are in
the unitary equivalence class of pure dephasing is still given by the result
derived above, i.e., by Eqs.~(\ref{eq:om2-fin}),(\ref{eq:om1-fin}). The
difference between the decoherence problems in this equivalence class lies
only in the initial values of $v_{x}(0)$, $v_{y}(0)$ and $v_{x}(0)$, which
enter into the explicit form of the control fields through the expressions~(%
\ref{eq:om2-fin}),(\ref{eq:om1-fin}). As a very simple example, consider the
case of transforming from the phase-flip channel to the unitarily equivalent 
$-$(phase-flip channel), i.e., $\sigma _{z}\mapsto -\sigma _{z}$. There
should, on physical grounds, be no essential difference between these two
cases. Indeed, in this case $U=\exp (i\pi \sigma _{y}/2)$, and the adjoint $%
\mathrm{SO}(3)$ matrix is a rotation by $\pi $ about the $y$-axis, i.e., $R=%
\mathrm{diag}\left( -1,1,-1\right) $. Thus $\vec{v}^{\prime }=R\vec{v}%
=(-v_{x},v_{y},-v_{z})$, and both the coherence $c=v_{x}^{2}+v_{y}^{2}$ and $%
v_{z}^{2}$ are invariant, so that the transformed control fields [compare to
Eqs.~(\ref{eq:om2-fin}),(\ref{eq:om1-fin})], given in terms of the
untransformed coherence vector, 
\begin{eqnarray}
\omega _{2}^{\prime }(t) &=&\left( \pm \right) \frac{+\gamma v_{x}(0)-\omega
_{0}(t)v_{y}(0)}{\sqrt{v_{z}^{2}(0)-2\gamma ct}}, \\
\omega _{1}^{\prime }(t) &=&\left( \pm \right) \frac{-\gamma v_{y}(0)-\omega
_{0}(t)v_{x}(0)}{\sqrt{v_{z}^{2}(0)-2\gamma ct}},
\end{eqnarray}%
have exactly the same divergence as before. The fact that there is a
difference at all, is a reflection of the fact that the mapping from $%
\mathrm{SU}(2)$ to $\mathrm{SO}(3)$ is homomorphic (double-valued)
\cite{Cornwell:84II}; indeed, the master equation~(\ref{eq:mastereq})
is invariant 
under the transformation $\sigma _{z}\mapsto -\sigma _{z}$, but this is not
the case in the coherence vector representation [the corresponding $M_{0}$
in Eq.~(\ref{eq:vdot})\ is not invariant].

It is worth emphasizing that the equivalence class of pure dephasing is the
entire group $\mathrm{SU}(2)$, but it excludes important processes such as
spontaneous emission (represented by $\sigma _{-}\notin \mathrm{SU}(2)$].
More generally, linear combinations of Pauli matrices, corresponding
to affine shifts in the coherence vector representation, are not in the
equivalence class of pure dephasing. The solution of the control problem for
such processes is deferred to a future publication; it involves solving the
non-linear tracking equation~(\ref{eq:track}) for these cases.

\section{Singularities}

\label{singularity}

It is insightful to reformulate the above control problem in terms of the tracking
control framework of Ref.~\cite{Zhu:99}, which allows one to study the
nature of the control field singularity. We once again linearize the
quadratic control objective $c(t)=c(0)$ into a two-dimensional form where we
wish to separately control $v_{x}(t)$ and $v_{y}(t)$ (in particular, keep
them constant), using the two control fields $\omega _{1}(t)$ and $\omega
_{2}(t)$. Following the notation of Ref.~\cite{Zhu:99} as closely as
possible we have [recall Eqs.~(\ref{eq:vdot1}),(\ref{eq:M(t)})]

\begin{equation}
\dot{\vec{v}}=(M_{0}+\sum_{j=0}^{2}\omega _{j}(t)\Lambda _{j})\vec{v}+\vec{k}%
,  \label{eq:vdot2}
\end{equation}%
and the control objectives $S_{1}=v_{x}$ and $S_{2}=v_{y}$ are 
\begin{equation}
S_{i}(t)=\vec{v}^{t}(t)\cdot O_{i}\cdot \vec{v}(t)~~[=S_{i}(0)],
\label{eq:Si}
\end{equation}%
with 
\begin{equation}
O_{1}=\left( 
\begin{array}{ccc}
1 & 0 & 0 \\ 
0 & 0 & 0 \\ 
0 & 0 & 0%
\end{array}%
\right) ,\quad O_{2}=\left( 
\begin{array}{ccc}
0 & 0 & 0 \\ 
0 & 1 & 0 \\ 
0 & 0 & 0%
\end{array}%
\right) .
\end{equation}%
The matrices $\Lambda _{j}$, defined in Eq.~(\ref{eq:lambdas}), satisfy%
\begin{equation}
\lbrack O_{1},\Lambda _{1}]=[O_{2},\Lambda _{2}]=0.
\end{equation}%
Using this fact, Eq.~(\ref{eq:vdot2}), and the observation $\Lambda
_{j}^{t}=-\Lambda _{j}$, it is simple to show by explicit differentiation of
Eq.~(\ref{eq:Si}) that 
\begin{eqnarray}
\omega _{1}(t) &=&\frac{\dot{S}_{2}+\omega _{0}(t)\vec{v}^{t}[\Lambda
_{0},O_{2}]\vec{v}-\vec{v}^{t}\{M_{0},O_{2}\}\vec{v}-\vec{k}^{t}O_{2}\vec{v}-%
\vec{v}^{t}O_{2}\vec{k}}{\vec{v}^{t}[\Lambda _{1},O_{2}]\vec{v}},
\label{eq:om1} \\
\omega _{2}(t) &=&\frac{\dot{S}_{1}+\omega _{0}(t)\vec{v}^{t}[\Lambda
_{0},O_{1}]\vec{v}-\vec{v}^{t}\{M_{0},O_{1}\}\vec{v}-\vec{k}^{t}O_{1}\vec{v}-%
\vec{v}^{t}O_{1}\vec{k}}{\vec{v}^{t}[\Lambda _{2},O_{1}]\vec{v}},
\label{eq:om2}
\end{eqnarray}%
where $\{,\}$ denotes the anti-commutator and its appearance in the
expression involving the decoherence matrix $M_{0}$, rather than a
commutator, is a manifestation of the associated damping. These equations
are equivalent to Eqs.~(\ref{om1}),(\ref{om2}), and can be further
simplified by using the constraint of constant coherence ($\dot{S}_{1,2}=0$%
), and the explicit forms of the various vectors and matrices appearing in
them.

Clearly, a singularity arises when the denominators in Eqs.~(\ref{eq:om1}),(%
\ref{eq:om2}) vanish: 
\begin{eqnarray}
\vec{v}^{t}(t)[\Lambda _{1},O_{2}]\vec{v}(t) &=&0,  \label{eq:zero1} \\
\vec{v}^{t}(t)[\Lambda _{2},O_{1}]\vec{v}(t) &=&0.  \label{eq:zero2}
\end{eqnarray}%
Ref.~\cite{Zhu:99} distinguishes between several types of singularities:\
(i) A trivial singularity is the case when the denominators are zero over a
continuous time domain, (ii)\ A non-trivial singularity is the case when the
denominators are zero at isolated points. A trivial singularity can be
removed by taking higher order time-derivatives of Eq.~(\ref{eq:Si}) under
the conditions~(\ref{eq:zero1}),(\ref{eq:zero2}), until the trivial
singularity is removed (i.e., a non-zero denominator is found). If
derivatives of all orders result in a trivial singularity, the system is
uncontrollable.

In our case, using 
\begin{equation}
\lbrack \Lambda _{1},O_{2}]=\left( 
\begin{array}{ccc}
0 & 0 & 0 \\ 
0 & 0 & 1 \\ 
0 & 1 & 0%
\end{array}%
\right) ,\quad \lbrack \Lambda _{2},O_{1}]=\left( 
\begin{array}{ccc}
0 & 0 & 1 \\ 
0 & 0 & 0 \\ 
1 & 0 & 0%
\end{array}%
\right) ,
\end{equation}%
we have 
\begin{eqnarray}
\vec{v}^{t}(t)[\Lambda _{1},O_{2}]\vec{v}(t) &=&2v_{y}(t)v_{z}(t), \\
\vec{v}^{t}(t)[\Lambda _{2},O_{1}]\vec{v}(t) &=&2v_{x}(t)v_{z}(t).
\end{eqnarray}%
Since our control fields keep $v_{x,y}$ fixed, the singularity arises at the
isolated point $v_{z}(t_{b})=0$. We are thus dealing with a non-trivial
singularity. There are now two subcases:\ a) The singularity is of the form $%
\omega _{j}(t)=\alpha /0$ with $\alpha \neq 0$; b) The singularity is of the
form $\omega _{j}(t)=0/0$. In case b) one can apply L'Hospital's rule and
(sometimes) overcome the singularity. In our case, the numerators in Eqs.~(%
\ref{eq:om1}),(\ref{eq:om2}) involve the decoherence matrix $M_{0}$ (and the
affine shift $\vec{k}$), while the denominators involve only the controls,
so that \emph{generically} one cannot expect a cancellation as in case b).
Ref.~\cite{Zhu:99} concludes that in case a) there is no solution for the
field and the system is uncontrollable at the singular point. This
conclusion agrees with our geometric interpretation of Sec.~\ref{analysis}.

\section{Conclusions and Open Questions}

\label{conclusions}

In this work we have considered the problem of controlling the coherence of
a single qubit under circumstances where none of the encoding or dynamical
decoupling methods recently developed in quantum information science apply.
Instead we have employed a version of tracking control, where a control
field is continuously adjusted in order to satisfy the objective of constant
coherence. This is possible up to a finite time, which depends on the
initial coherence, at which the control field diverges.

There are various open questions suggested by these results.

i) While our original goal was to track the coherence $c=v_{x}^{2}+v_{y}^{2}$%
, we in fact solved the more restrictive problem of separately controlling $%
v_{x}$ and $v_{y}$. It would be interesting to consider the case where these
components are allowed to vary while truly trying to fix only $c$. In
particular, it would be interesting to see if this enables the extension of
the breakdown time.

ii) Expansion of the Hilbert space by including additional levels:
Controllability could improve if instead of having a two-level system we
were to use an $n$-level system, where we coherently control all the $n$
levels, but use just two for the qubit. Within this larger Hilbert space it
is possible that interference effects could be used profitably to maintain
the coherence of the two qubit levels, as e.g., in electromagnetically
induced transparency \cite{Harris:97}, or control of vibrational wavepackets 
\cite{Brif:01}.

iii) Periodic or continuous update of the control objective so as to reduce
the desired value of coherence: In this manner the singularity of the control
fields can be avoided for arbitrarily long times. It would be interesting to
formulate this as an optimal control problem, with the objective being,
e.g., the time-integral of coherence.

iv) As discussed above, the control fields depend crucially on the knowledge of the
initial state whose coherence we wish to maintain. For certain states [with $%
v_{z}(0)=0$] no such control is possible. Additionally we need to know $%
v_{x}(0)$ and $v_{y}(0)$ in order to correctly apply the control fields. The
full tracking control problem requires knowing even more: the full coherence
vector $\vec{v}(t)$, an impossible task due to the non-commutativity of the
observables involved. In general, one can envision performing quantum
tracking control based on the incomplete information gleaned in real time
from measuring an as large as possible set of \emph{commuting} observables.
It is an open problem to estimate the quality of the (partial) tracking one
can thus attain.

v) As mentioned in Sec.~\ref{purity}, the purity for non-unital decoherence
channels can actually increase without control. Unitary, open-loop control
in this case was studied in Ref.~\cite{Recht:02}, It would be interesting to
explore coherence tracking control for this class of channels.

vi) Finally, an important extension of the results reported here would
be to problems involving more than one qubit, e.g., in order to
preserve entanglement.

\begin{acknowledgments}
Financial support from the DARPA-QuIST program (managed by AFOSR under
agreement No. F49620-01-1-0468), the Sloan Foundation, and PREA (to D.A.L.)
is gratefully acknowledged.
\end{acknowledgments}


\end{document}